\documentclass[sigconf]{acmart}
\usepackage{placeins}

\AtBeginDocument{%
  }

\graphicspath{{figures/}}

\renewcommand{\footnotetextcopyrightpermission}[1]{}

\begin{document}

\title{HBF-EventLab: A Discrete-Event Simulator for Studying
Host-Device Coordination in High Bandwidth Flash}


\author{Ziqing Mei}
\authornote{These authors contributed equally to this work.}
\affiliation{%
  \institution{Shanghai Jiao Tong University}
  \city{Shanghai}
  \country{China}}
\email{mei\_ziqing@sjtu.edu.cn}

\author{Yiwei Huang}
\authornotemark[1]
\affiliation{%
  \institution{Shanghai Jiao Tong University}
  \city{Shanghai}
  \country{China}}
\email{hyw037@sjtu.edu.cn}

\author{Zhiyi Zeng}
\authornotemark[1]
\affiliation{%
  \institution{Huazhong University of Science and Technology}
  \city{Wuhan}
  \country{China}}
\email{zhiyi@hust.edu.cn}

\author{Yuhao Xiao}
\authornotemark[1]
\affiliation{%
  \institution{China University of Geosciences}
  \city{Wuhan}
  \country{China}}
\email{x\_rainh@163.com}

\author{Pu Pang}
\affiliation{%
  \institution{Shanghai Jiao Tong University}
  \city{Shanghai}
  \country{China}
}
\email{pangpu@cs.sjtu.edu.cn}

\author{Quan Chen}
\affiliation{%
  \institution{Shanghai Jiao Tong University}
  \city{Shanghai}
  \country{China}
}
\email{chen-quan@cs.sjtu.edu.cn}

\author{Deze Zeng}
\affiliation{
  \institution{China University of Geosciences}
  \city{Wuhan}
  \country{China}
}
\email{deze@cug.edu.cn}

\author{Minyi Guo}
\affiliation{%
  \institution{Shanghai Jiao Tong University}
  \city{Shanghai}
  \country{China}
}
\email{guo-my@cs.sjtu.edu.cn}

\renewcommand{\shortauthors}{Mei et al.}

\begin{abstract}
  High Bandwidth Flash (HBF) combines NAND flash density with high-bandwidth accelerator connectivity, offering additional memory capacity for large language model inference. However, limited hardware availability makes it difficult to evaluate how host behavior and device resource constraints jointly affect HBF request processing. We present HBF-EventLab, an open-source, transaction-level discrete-event simulator based on the Open Compute Project HBF specification. HBF-EventLab connects host request dispatch, command admission, fragmented write assembly, read dispatch, NAND execution, and completion feedback within a shared event loop. It explicitly models distinct resource lifetimes and separates configurable scheduling policies from ordering and resource constraints, enabling controlled exploration of host–device coordination. 
  Experiments show how resource limits, address interleaving, fragment arrival patterns, and scheduling policies affect write completion and request latency. 
  The results highlight the need to consider resource limits and scheduling policies jointly, accounting for their interactions with write assembly and read eligibility. We also integrate HBF-EventLab with real GPU execution of Qwen3-0.6B, combining profiled GPU activity durations with simulated weight-access delays to estimate time to first token and time per output token, with logits and output tokens remaining bitwise identical to those from native GPU execution. HBF-EventLab provides a configurable foundation for studying HBF mechanisms, exploring scheduling policies, and assessing how HBF access delays affect application-level timing. 
  {\bf The source code is publicly available at}
\href{https://github.com/sjtu-epcc/HBF-EventLab}{%
  \textcolor{blue}{\nolinkurl{https://github.com/sjtu-epcc/HBF-EventLab}}}.

\end{abstract}

\maketitle

\fancyhead[LE,RO,RE]{}

\section{Introduction}
\label{sec:introduction}

The rapid development of large language models (LLMs) has placed increasing demands on memory capacity, driven by growing model sizes and the need to support longer contexts. 
In addition to model weights, inference systems retain a KV cache whose size grows with context length and batch size~\cite{infinigen2024}.
For large models and long-context workloads, the combined footprint of model weights and the KV cache can exceed the high-bandwidth memory (HBM) capacity available on a GPU~\cite{ren2025offloading}. To accommodate these workloads, existing inference systems offload a portion of the model weights and KV cache to host memory or solid-state drives (SSDs), transferring the required data to the GPU as inference proceeds~\cite{flexgen2023,infinigen2024,ren2025offloading}. This approach expands the capacity available to inference, while making data movement between the GPU and these additional memory and storage tiers an important factor in performance~\cite{flexgen2023,infinigen2024}.

High Bandwidth Flash (HBF), an emerging memory technology based on stacked NAND flash, offers a potential way to alleviate this data movement overhead~\cite{sandisk2025hbf}. In conventional offloading systems, host memory and SSDs reside outside the GPU package, and accessing offloaded data typically requires transfers over interconnects such as PCIe. The limited bandwidth of these links can make fetching model weights and KV cache a bottleneck during inference~\cite{flexgen2023,infinigen2024,flexinfer2025}. HBF addresses this limitation by placing flash storage close to the GPU and combining parallel NAND access with high-bandwidth UCIe connectivity~\cite{skhynix2026hbf}. Together with the high storage density of NAND flash, this design enables HBF to complement HBM with greater capacity and high read bandwidth for inference workloads~\cite{sandisk2025hbf}.

Despite its promise, commercial HBF hardware is not yet widely available, limiting opportunities for direct experimental evaluation~\cite{sandisk2025hbfroadmap}. Although the Open Compute Project (OCP) specification defines HBF's architecture, interfaces, and operational constraints, substantial design space remains in request scheduling, data placement, and coordination between the host and the device~\cite{ocp2026hbf}. Exploring these choices and their performance requires a configurable simulator. Existing DRAM, NAND, and SSD simulators do not directly capture HBF's distinct interface, NAND organization, and controller behavior~\cite{ramulator2024,nandflashsim2012,mqsim2018,simplessd2018}. Therefore, we develop an HBF-specific simulator to model host-device coordination and NAND operations based on the OCP specification~\cite{ocp2026hbf}.

A well-designed HBF simulator should faithfully model the device organization, resource constraints, and operational behavior defined by the OCP specification, while remaining extensible where implementation choices are left open~\cite{ocp2026hbf}. Resource configurations across dies, channels, and banks, together with host request dispatch, device scheduling, and request arrival patterns, can affect bandwidth utilization and access latency, thereby influencing LLM inference performance. 
The simulator should provide configurable resource parameters and extensible host and device policies, enabling researchers to evaluate these interactions and explore alternative designs.


To this end, we develop HBF-EventLab, a transaction-level discrete-event simulator for studying HBF host--device coordination. HBF-EventLab follows the OCP Specification~\cite{ocp2026hbf}.
It connects a finite host workload to device admission, assembly, execution, completion, and response feedback within one event loop. Policy interfaces separate selection decisions from the legality and ownership constraints that every decision must preserve. The observation model distinguishes original logical work, individual command attempts, and assembly-context generations, allowing experiments to relate completion outcomes and latency to the underlying request and resource lifetimes.


HBF-EventLab is open source\footnote{\url{https://github.com/sjtu-epcc/HBF-EventLab}} and allows researchers to configure host dispatch and device scheduling policies to study their effects on request latency and completion.
It can also be integrated with real GPU inference through an adapter to the llama.cpp/ggml CUDA executor~\cite{llamacpp}.
Our experiments show that varying resource configurations and scheduling policies produces the expected effects on request latency and completion (Section~\ref{sec:evaluation-backend}).
Further integration with real GPUs enables researchers to estimate time to first token (TTFT) and time per output token (TPOT) for LLM inference, as demonstrated with Qwen3-0.6B~\cite{qwen3_06b} on an NVIDIA GeForce RTX 5070 in Section~\ref{sec:gpu-integration}.


This paper makes three contributions:
\begin{itemize}
    \item An executable HBF simulator that connects
    host behavior, finite device resources, and response-driven
    retry and recovery within a closed request--response loop.

    \item Policy and observation interfaces that support
    experiments under shared execution constraints and preserve
    the distinctions among logical work, command attempts,
    and assembly generations.

    \item Comprehensive experimental results showing how
    resource distribution, fragment arrival, and scheduling
    opportunities affect assembly progress, completion
    outcomes, and read/write latency.
\end{itemize}




\section{Background and Related Work}
\label{sec:background}


This section introduces necessary concepts for HBF as well as the related work.

\subsection{HBF Architecture and Data Organization}

{\bf Physical Architecture.}
HBF couples NAND capacity with the accelerator-facing bandwidth through multiple independent UCIe channels, combining wide interfaces with NAND parallelism. OCP HBF v0.7.0~\cite{ocp2026hbf} describes a Base Die and a Core Die stack connected through TSVs. The Base Die handles host communication, command processing, and NAND data transfers. Up to 16 independent channels provide full-duplex interfaces to corresponding portions of the device address space.

\begin{figure}[t]
  \centering
  \includegraphics[width=0.6\linewidth]{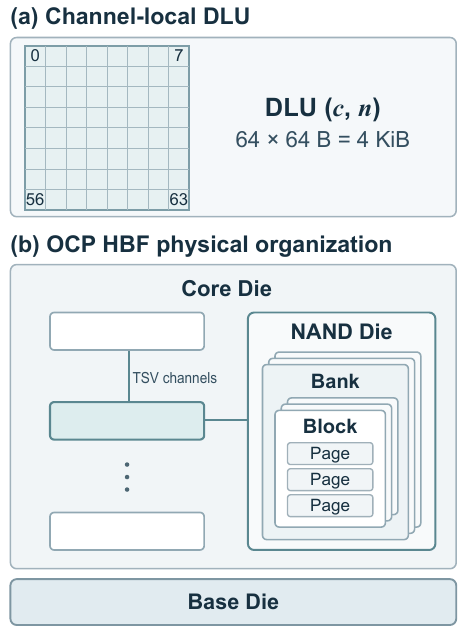}
  \caption{Schematic OCP HBF physical organization and channel-local DLU granularity. 
  In this paper, a \emph{channel} denotes a UCIe Host Channel, a \emph{die} denotes an individual NAND Die, and a \emph{fragment} denotes a 64-byte unit. \emph{Core Die} refers to the NAND stack.
  }
  \Description{The figure contains two panels. Panel (a) shows a channel-local 4-KiB DLU as an 8-by-8 grid of 64-byte Fragments. Panel (b) illustrates the HBF physical organization, including the Base Die and the Core Die stack connected through TSV channels. An individual NAND Die is expanded to show the hierarchy of Banks, Blocks, and Pages. The logical DLU organization and physical NAND hierarchy are distinct views, and no fixed one-to-one mapping between them is implied.}
  \vspace{-10pt}
  \label{fig:hbf-organization}
\end{figure}

{\bf NAND Physical Organization.}
Core Die Channels connect the Base Die to the NAND dies, distinct from host-facing UCIe channels. The NAND hierarchy comprises Core Die Channels, dies, Banks, Blocks, and Pages (Figure~\ref{fig:hbf-organization}(b)). Reads and programs operate on Pages, whereas erases operate on Blocks. Independent Banks provide operation-level parallelism; the degree of parallelism achieved depends on address placement and scheduling. The Base Die also provides Bank Caches (Cache Buffers), with a default configuration of two 4-KiB slots per Bank holding up to two Page copies, separate from write-assembly storage.

{\bf Logical Data Organization.}
A channel-local Device Logical Unit (DLU) contains 4 KiB, or 64 consecutive fragments (as shown in Figure~\ref{fig:hbf-organization}(a)). Host software interleaves Global addresses across channels; the Base Die maps local addresses to NAND resources. A DLU is a logical unit whose physical placement depends on the Base Die architecture. Section~\ref{subsec:design-dlu} defines the reference mapping used in HBF-EventLab.

{\bf Wear Leveling and Zone Remapping.}
A zone is an equal-sized region of a channel's address space used for remapping. HBF supports vendor-specific Base-die-driven wear leveling and host-controlled zone remapping without device-side migration of live data. Before a zone swap, the Host must invalidate both zones and stop reads to them; it is responsible for subsequent data rewriting~\cite{ocp2026hbf}.

\subsection{OCP HBF Interface and Request Processing}
\label{subsec:ocp-interface}

{\bf Host Interface and Command Model.}
HBF uses an AXI-based interface over UCIe, with separate request and response channels. AXI IDs associate commands with responses: responses sharing an ID remain ordered, while different IDs permit out-of-order completion. We focus on NAND-backed Flash I/O; Scratchpad and CSR/ADMIN transactions are outside the modeled scope.

The OCP defines Maximum Outstanding Commands Supported and Maximum Outstanding Commands Selected (both abbreviated MOCS), with $MOCS_{\mathrm{selected}}\le MOCS_{\mathrm{supported}}$. The Host-selected limit caps outstanding commands. In our model, a Pending DLU is a live write context spanning collection, readiness, and programming until retirement; the Pending DLU limit bounds these contexts. The simulator sets these limits through configuration rather than CSR/ADMIN transactions.

{\bf Read Request Processing.}
Host reads have a 64-byte granularity; supported bursts can span up to 4 KiB without crossing a Page boundary. Regular reads may accumulate before dispatch, while a Batch Read Hint permits dispatch without waiting for the accumulation timer. Section~\ref{subsec:design-read} defines the batch-boundary interpretation used by HBF-EventLab.

Cache hits can be served directly when ordering permits. Misses require NAND Sense operations, which preserve same-Bank ordering while allowing independent Banks to operate concurrently.

{\bf Write Request Processing.}
Host writes are non-posted: each command receives one completion response, with success reported after programming. The Base Die assembles incoming fragments into complete 4-KiB DLUs before programming. At the Pending limit, existing contexts can still receive missing fragments, while commands opening new DLUs receive a Pending-limit error. A configurable deadline starts with the first fragment; incomplete assembly produces errors and may require the Host to resend the DLU.

Within each Block, Pages must be written sequentially without skipping, while independent Blocks can progress concurrently. A legal Page-0 write initiates device-managed Block Erase; ordering violations produce errors.

{\bf Read/Write Ordering and Error Handling.}
Reads and writes to the same fragment retain their required order even across AXI IDs. A read following a Pending write under the ordering rules receives the collected fragment if available, or status 0xA if that fragment is missing. Read and write response fields convey status and additional error information for Host recovery, including address, capacity, timeout, and write-order errors.

\subsection{Related Work}
\label{subsec:related-work}

{\bf HBF Architectures and Applications.}
Recent studies have explored the integration of High Bandwidth Flash (HBF) into accelerator memory hierarchies, primarily focusing on architectural integration and workload-specific optimizations. H$^3$ proposes a hybrid HBM--HBF architecture that places read-only data in HBF while retaining other data in HBM, targeting capacity-intensive LLM inference workloads~\cite{h3_2026}. Mem-Village explores an HBF--HBM--GPU architecture integrated through a glass interposer, focusing on package-level integration and layer-wise weight streaming~\cite{memvillage2025}. Son et al. systematically investigate the opportunities and challenges of using HBF as a major memory component for LLM inference, highlighting the trade-offs among capacity, read bandwidth, write performance, and endurance~\cite{hbfopportunities2026}.

Other work targets weight streaming~\cite{flint2026}, KV-cache placement and heterogeneous memory management~\cite{hbfkv2026,hbfhma2026,hbfhotcold2026,sparsetier2026}, and read-reclaim endurance~\cite{readreclaim2026}. These studies motivate HBF placement and access policies at the architecture and application levels. Complementing this work, HBF-EventLab provides a configurable backend for examining how the resulting access patterns interact with command credits, write assembly, and device scheduling.

{\bf Flash Storage Systems and Simulation.}
Open-channel interfaces and flash-resource partitioning expose placement, scheduling, and isolation opportunities to host software~\cite{lightnvm2017,flashblox2017,flashparallelism2016}. SSD simulators and emulators model flash resources, firmware, and host interactions~\cite{mqsim2018,simplessd2018,femu2018,amber2018}, but target conventional SSD interfaces and management mechanisms. HBF-EventLab shares the emphasis on explicit flash resources and host–device interaction, while targeting the OCP HBF transaction model. It represents fragment assembly, Pending-context occupancy, and command-credit lifetimes separately, enabling studies of admission and assembly constraints beyond conventional SSD command processing.

{\bf Memory and GPU Simulation.}
DRAM simulators provide configurable resource and timing models~\cite{ramulator2016,dramsim3_2020}. Ramulator 2.0 separates component interfaces from implementations, supporting extensible policies and memory specifications~\cite{ramulator2024}; this modularity informs our design.
GPGPU-Sim and Accel-Sim model GPU execution and memory-system behavior~\cite{gpgpusim2009,accelsim2020}. Vidur, LLMServingSim, and LLMCompass study LLM performance through profiling, hardware models, and scheduling exploration~\cite{vidur2024,llmservingsim2024,llmcompass2024}. HBF-EventLab complements these execution and system-level models with an OCP-based HBF backend. Its incremental request interface and completion feedback provide a coupling boundary through which external execution models can submit memory accesses and incorporate HBF service delays.

\begin{figure*}[t]
  \centering
  \includegraphics[width=0.8\textwidth]{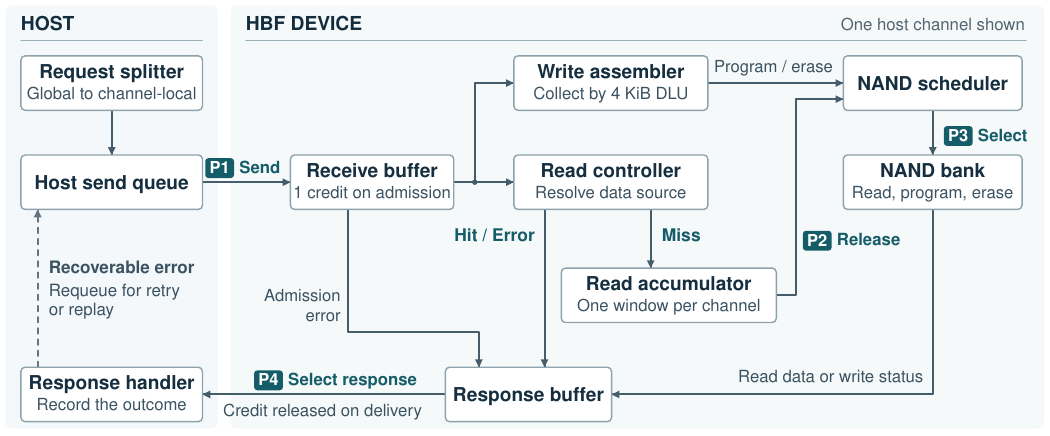}
  \caption{Request and response paths for one host channel in HBF-EventLab.}
  \Description{The host splits global requests into channel-local
  commands and sends them to the device. The device assembles write
  fragments for NAND programming. Read hits and errors go directly to
  the response buffer. Read misses enter a per-channel window before
  NAND scheduling. Delivering a response to the host releases one
  command credit. P1 selects the next command to send, P2 releases
  accumulated reads, P3 selects the next NAND operation, and P4 selects
  the next response to return. The dashed arrow shows the retry or
  replay path after a recoverable error.}
  \label{fig:design-architecture}
\end{figure*}

\section{HBF-EventLab Design}\label{sec:design}

This section describes our proposed HBF simulator, HBF-EventLab.
Section~\ref{subsec:design-architecture} presents the simulator's overview and modeling scope.
Section~\ref{subsec:design-dlu} explains address mapping and the NAND resource model.
Section~\ref{subsec:design-write} describes write admission, DLU assembly, resource
lifetimes, and failure recovery.
Section~\ref{subsec:design-read} introduces read processing, including Pending-write
forwarding, cache access, and read-window dispatch.
Section~\ref{sec:scheduling} presents the scheduling policies for Host sending,
read-window release, NAND execution, and response arbitration.
Finally, Section~\ref{sec:event-execution} explains event execution, causal ordering,
and runtime observability.


\subsection{Overview}\label{subsec:design-architecture}

Figure~\ref{fig:design-architecture} shows the request and response paths for one channel. The Host tracks request release times and Global byte addresses, splits requests into channel-local commands, and collects completion feedback. The Base Die models command processing and responses; the Core Die supplies NAND execution resources.

The Host selects commands through the Host Sending Policy
(\textbf{P1}), which chooses the next eligible command from the
per-channel send queue. A successful handoff transfers ownership
of the command to the Device, reserves receive-buffer capacity,
and initiates request-link transmission. After the command arrives
at the Device, formal acceptance releases the ingress reservation
and allocates one command credit on its channel. The credit remains
occupied until the corresponding response is delivered to the Host.
Device acceptance and subsequent operation-specific admission
are distinct: for example, an accepted write command may still
return an error during write-assembly checks without contributing
data to an assembly context. A handoff not taken consumes no command credit; after formal acceptance, even an error response retains its credit until delivery.

Reads are served through Pending forwarding, Bank Cache, or windowed NAND access. {\bf P2} releases accumulated misses, {\bf P3} selects eligible NAND operations, and {\bf P4} arbitrates responses. 
Writes assemble into DLUs before programming. Figure~\ref{fig:design-architecture} shows functional modeling components rather than a prescribed hardware partition.

The event engine coordinates transmission, acceptance, assembly, NAND operations, and completion feedback. Each channel has independent request and response paths with serialization, propagation delay, and finite buffers. Core Die Channels provide ideal routing without additional contention or delay; NAND service times remain fixed during a run.

The modeled OCP HBF v0.7.0 subset~\cite{ocp2026hbf} represents transactions rather than AXI handshakes or UCIe flits, and excludes cell-level behavior and ECC. Timing parameters support controlled mechanism comparisons and have not been calibrated against physical HBF hardware. GPU integration is evaluated separately.

\subsection{Address Mapping and NAND Resource Model}
\label{subsec:design-dlu}

The model separates logical assembly units from NAND placement and execution resources~\cite{ssdtradeoffs2008,flashparallelism2016}. A Pending DLU context tracks received fragments for one DLU. The reference layout maps each DLU to one 4-KiB Page; fragmented writes supply separate fragments, while an enabled 4-KiB burst supplies the complete DLU.

\begin{figure*}[t]
  \centering
  \includegraphics[width=0.8\textwidth]
    {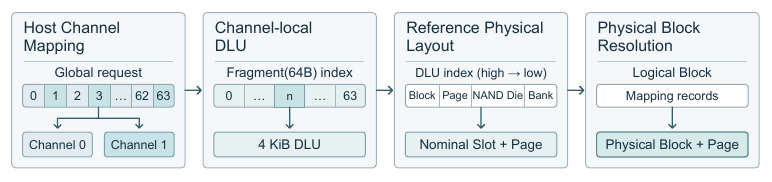}
  \caption{Four-stage address translation and physical-block
  resolution in HBF-EventLab. Global requests are interleaved
  across UCIe Host Channels, mapped to channel-local DLUs,
  decoded by \texttt{ReferenceAddressLayout} into nominal
  NAND locations, and resolved through mapping records
  to their current physical Blocks.}
  \Description{The figure illustrates four stages of address
  translation. First, a Global request is interleaved across
  example Host Channels. Second, each local byte address
  identifies a channel-local 4-KiB DLU and one of its
  64-byte fragment positions. Third, ReferenceAddressLayout
  decodes the DLU index into a nominal NAND location and
  Page using the Bank, NAND Die, Page, and Block dimensions.
  Finally, mapping records resolve a Logical Block to its
  current Physical Block while preserving the Page position.}
  \label{fig:address-mapping}
\end{figure*}

Figure~\ref{fig:address-mapping} summarizes the four stages
of address translation in HBF-EventLab: Global requests are
interleaved across Host Channels, converted into channel-local
DLUs, decoded into nominal NAND locations, and finally
resolved to physical Blocks through mapping records.
In particular, a 4-KiB Global request need not form one
local DLU: channel interleaving can distribute its fragments
across several channels and local DLUs. Conversely, fragments
originating from different Global requests can contribute
to the same channel-local DLU.

For a Global byte address $a$, let $a_0$ be the configured HBF Global base address shared by all requests and $x=a-a_0$ its offset from that base. Let $C$
denote the number of Host Channels and $I$ the interleaving
stripe size in bytes. The round-robin interleaver computes
the destination channel $c$ and local byte address $\ell$ as
\begin{equation}
  c=\lfloor x/I\rfloor\bmod C,\qquad
  \ell=I\lfloor x/(CI)\rfloor+(x\bmod I).
  \label{eq:design-interleave}
\end{equation}

The Host frontend applies this mapping to workload ranges
and splits them into channel-local operations. The local
DLU index is $n=\lfloor\ell/4096\rfloor$, and the fragment
index $f=\lfloor(\ell\bmod4096)/64\rfloor$ identifies one
of the 64 positions within that DLU. The implementation
uses interleaving granularities supported by the OCP HBF
configuration. These operations correspond to the first
two stages of Figure~\ref{fig:address-mapping}.

Our reference layout stripes DLUs across Banks and dies before advancing Pages and Blocks. With $N_p$ Pages per Block, $N_d$ dies per channel, and $N_b$ Banks per die, the same channel-local DLU index $n$ is encoded as
\begin{equation}
  n=((bN_p+p)N_d+d)N_b+\beta,
  \label{eq:design-layout}
\end{equation}
where $b$, $p$, $d$, and $\beta$ denote Block, Page, die, and Bank indices. The decoder and encoder provide consistent conversions between this index and nominal NAND coordinates.

HBF-EventLab distinguishes three levels of address identity:
the Logical Block, the Nominal Slot, and the Physical Block.
The Logical Block identifies the Host-visible address and
is used to maintain write-order and retirement state.
The Nominal Slot represents the Block's default location
in the reference layout, while the Physical Block identifies
the actual media resource after address translation.
This separation allows physical Block replacement without
changing the logical address or its associated ordering
state. During address resolution, the simulator first checks
whether the Logical Block has been retired, then resolves
its nominal location through the configured mapping and
replacement records to determine the current Physical Block.
The final stage of Figure~\ref{fig:address-mapping}
illustrates this resolution from a Logical Block through
mapping records to its current Physical Block and Page.

Spare Blocks are excluded from Host-addressable capacity. The reference replacement policy selects the next healthy spare within the same Bank, preserving page offsets; spare exhaustion permanently retires the Logical Block.

An internal API supports channel-local Zone Mapping. The CSR/ADMIN path is not implemented, and zones must align to complete Block Groups (all Banks' Blocks sharing a Block index within a channel), a stricter constraint than OCP's NAND-Block-size alignment. Each channel connects to one ideal Core Die Channel in the reference topology.

A physical Bank is the primary unit of NAND
execution-resource management. As a conservative
resource baseline, each Bank executes at most one
Read (Sense), Program, or Erase operation at a time,
while independent Banks can execute operations
concurrently subject to other modeled resources.
The model does not impose an additional die-wide lock
for the full duration of a Bank operation. The Base Die
also maintains two 4-KiB Cache Buffer slots per Bank,
matching the configured Bank Cache count.
These Cache Buffers are modeled as finite resources
separate from the per-channel Pending write-assembly
storage.

Each programmed Page is stored sparsely as 64 symbolic DataTokens. Each token identifies the content of one fragment rather than storing its bytes, enabling fragment-coverage and read-after-write identity checks. Unprogrammed Pages are treated as erased.

The media model also maintains the programming progress
and program/erase count (PEC) of each physical Block.
A successful Erase clears the Block's stored Page data,
resets its next-page programming position, and increments
its PEC. The PEC is associated with the physical Block:
Block replacement or Zone Remapping changes the
logical-to-physical association but does not transfer
or reset the erase history of an existing physical Block.
These counters provide functional wear-state bookkeeping
rather than a calibrated NAND endurance model.

\subsection{Write Processing and Resource Lifetimes}
\label{subsec:design-write}

{\bf Write Admission and DLU Assembly.}
HBF-EventLab separates write-command admission, DLU assembly, and NAND programming into distinct stages. After a command is accepted at the device interface, the write assembler checks its address, fragment overlap, block write order, and available assembly capacity before modifying the corresponding DLU context. An accepted command may still be rejected during assembly admission; such a rejection does not partially modify the context. Each context maintains a 64-bit received bitmap, symbolic fragment data, an assembly-generation identifier, and the associated command state. A DLU becomes Ready when all 64 distinct 64-byte fragments have been accepted. To avoid partial updates, the simulator checks all fragments in a command for overlap before committing any of them.

After request-link transfer and admission checks, a complete 4-KiB burst populates all fragment positions atomically at the acceptance timestamp. Link serialization accounts for its transfer time; no fragmented-write assembly deadline is needed. For fragmented writes, the deadline starts with context creation and applies only while fragments are missing. Ready and Programming DLUs do not incur assembly timeouts. An assembly generation identifies a particular context instance, preventing stale timeout events from affecting its replacement. It is distinct from the logical content version, which can persist across retries and replay; same-Tick deadline ordering follows Section~\ref{sec:event-execution}.

{\bf Write Ordering and Auto Erase.}
Write ordering is maintained independently for each Logical Block. HBF-EventLab uses separate progress cursors for fragment reception and NAND programming. For reception, the simulator interprets the OCP requirement that a prior page be sent as requiring all fragments of that Page to be accepted before assembly of the next Page can begin. This is the model's interpretation of the reception boundary, which is not precisely defined by the specification. For programming, however, the next Page cannot begin Program until the preceding Page has been programmed successfully. This separation allows reception of Page $p+1$ to overlap with the programming of Page $p$, while preserving program order. Different Logical Blocks maintain independent ordering state and can progress concurrently subject to shared physical resources. The implementation maintains at most one Collecting Page and one Programming Page per Logical Block.

Consistent with the OCP page-zero restart behavior, a legal write beginning at Page~0 initiates one Auto Erase for the corresponding Block. In the simulator, acceptance of the first fragment for Page~0 registers the erase operation, which can proceed while the remaining fragments are being assembled. No Page in the Block can be programmed until the erase completes, and subsequent fragments do not trigger additional erase operations. Before admitting a new Page~0 start, the model waits for prior writes, erase operations, outstanding responses, logical read dependencies, and active physical mapping leases on the target to drain. A lease protects the physical target used by an active media operation against unsafe replacement or reuse. These dependencies prevent rewriting content or targets still in use. The erase and restart behavior follows OCP Sections~5.4.1 and~11.5.2.2~\cite{ocp2026hbf}, while the detailed admission and preparation states are implementation mechanisms.

{\bf NAND Programming and Resource Lifetimes.}
A complete DLU becomes a Program candidate only when its assembly is complete, the corresponding Block has completed erase preparation, and the per-Block program-order constraint permits execution. The NAND Scheduler (P3) selects among eligible operations when the target Bank becomes available (Section~\ref{sec:scheduling}). Once issued, Program runs to completion before the Bank is released. A Ready DLU continues to occupy its Pending slot while waiting for a Bank and throughout the Program operation. On successful completion, the simulator commits the PageData to the modeled NAND media state, advances the program-order cursor, retires the assembly context, and releases its Pending slot. The corresponding command responses then become eligible for delivery.

The write path maintains three independent per-channel resource limits, summarized in Table~\ref{tab:write-resource-lifetimes}. A Host-open DLU is tracked from its first successful write handoff until all handed-off commands for it receive responses, limiting Host concurrency. Unlike this Host-side record, a Pending context tracks Device assembly and programming; command credits track individual accepted commands. Their distinct lifetimes require separate accounting.

\begin{table}[t]
\centering
\caption{Three per-channel limits and their accounting boundaries.}
\label{tab:write-resource-lifetimes}
\small
\setlength{\tabcolsep}{3pt}
\begin{tabular}{@{}p{0.18\columnwidth}
                  p{0.27\columnwidth}
                  p{0.45\columnwidth}@{}}
\hline
\textbf{Limit} &
\textbf{Where tracked} &
\textbf{What occupies one slot} \\
\hline
Host-open DLUs &
Host send queue and response handler &
One DLU from its first successful write handoff until all handed-off writes for it have received responses. \\
\hline
Command credits ($M$) &
Device admission and response delivery &
One accepted read or write command until its own response reaches the Host. \\
\hline
Pending DLUs &
Write assembler &
One DLU assembly context from the first fragment until the context is retired. \\
\hline
\end{tabular}
\end{table}

\begin{figure}[t]
  \centering
  \includegraphics[width=0.85\columnwidth]
    {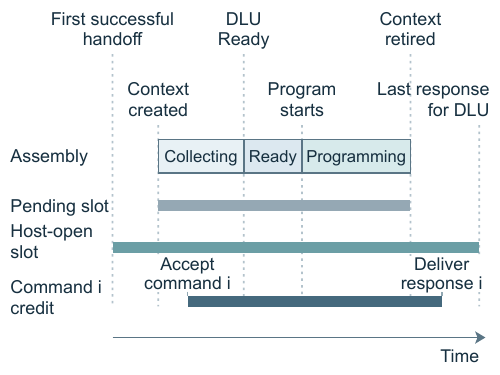}
  \caption{Resource lifetimes for one successful fragmented DLU
  write. The Pending slot spans context creation to retirement,
  the Host-open slot spans the first successful handoff to the
  last response for the DLU, and each command credit spans its
  own acceptance to response delivery.}
  \Description{Time progresses from left to right. The assembly
  context progresses through Collecting, Ready, and Programming.
  The Pending slot is occupied from context creation until context
  retirement. The Host-open slot remains occupied from the first
  successful handoff for the DLU until its last response is
  delivered. The credit for command i is occupied from device
  acceptance of command i until delivery of its corresponding
  response.}
  \label{fig:design-lifetimes}
\end{figure}

These limits are independent. Multiple accepted fragments can consume separate command credits while sharing one Host-open slot and one Pending slot. When Pending capacity is exhausted, a new DLU may receive a Pending-limit error even if command credits remain available. A valid fragment for an existing assembly context can still be accepted. A Pending-limit rejection can be retried by the Host while the retained fragments remain useful for making progress. Command credits are shared by reads and writes.
Figure~\ref{fig:design-lifetimes} illustrates these overlapping resource lifetimes.

{\bf Failure and Recovery.}
Write failures are handled according to the stage at which they occur. An assembly-level rejection, such as a Pending-limit or overlap error, does not partially modify an existing context. When an assembly deadline expires before all fragments arrive, the incomplete context is retired and error responses are returned for its associated commands. The Host waits for previously handed-off attempts for that Page to drain, reconstructs the complete Page from retained input data, and resubmits it using ordinary Write commands. An assembly timeout therefore requires retrying the same Page rather than automatically replaying the entire Block. For a Page~0 retry targeting the same Block and content version, existing erase preparation can be reused instead of issuing another erase.

A Program failure terminates the failed DLU and any already accepted subsequent DLUs in the same Logical Block. The failed DLU receives a Program Failure status (0x7), while affected subsequent DLUs receive the corresponding sequence error (0x6, info=2). The injected Program outcome specifies whether the physical Block remains reusable or becomes unavailable; an unusable outcome marks the physical Block unavailable and can invoke the spare-replacement mechanism described in Section~\ref{subsec:design-dlu}.

After old attempts drain, the Host replays retained content from Page~0 through the failed Page using ordinary writes. Restarting at Page~0 establishes the erase and sequential-programming process for the recovered content. This endpoint follows our interpretation of OCP Section~11.5.2.7~\cite{ocp2026hbf}. A failure during replay retains the furthest required endpoint. Recovery ends when programming reaches that endpoint; missing retained data or exhaustion of the replay-pass budget terminates the affected logical work.

\subsection{Read Processing and Dispatch}
\label{subsec:design-read}

{\bf Pending Write Forwarding and Cache Lookup.}
When processing a Read Command, HBF-EventLab first checks
whether the target DLU has an active Pending DLU context applicable under read/write ordering. If the
requested fragment has already been received, the device
forwards the corresponding DataToken directly. Otherwise,
it returns a missing-fragment status (0xA), without falling
back to potentially stale data in NAND or a Cache Buffer. This
behavior follows the Pending read semantics defined in
Section 5.4.3 of the OCP specification~\cite{ocp2026hbf}.
The current implementation supports unordered 64-byte
Pending reads; ordered Pending reads remain outside the
supported subset and are reported as a model-capability
diagnostic.

If no applicable Pending DLU context exists, the device checks
the target Bank's Cache Buffers for a valid copy of the requested
page. An initial Cache hit satisfying ordering constraints incurs a configurable cache-access latency (100\,ps by default) and bypasses both the read window and NAND sensing. On a Cache miss, the request
proceeds to the read-candidate management and dispatch
process. Waiting reads are rechecked for Cache hits, including before NAND issue. An initially missed read can therefore leave the window before release if its page becomes cached and ordering permits service.

{\bf Read Candidate and Issue.}
Accepted reads retain logical content dependencies while awaiting dispatch, without reserving a Bank or binding a physical target. At issue, the device rechecks mapping, ordering, Cache state, and Bank availability, then binds the target and samples symbolic data. Completion releases the Bank; the result remains protected in its Cache slot until ownership transfers to bounded return storage.

{\bf Read Window.}
HBF-EventLab maintains one read window per
UCIe Host Channel. A window can collect reads targeting
multiple NAND Dies and Banks within the channel, allowing
subsequent NAND scheduling to exploit available
parallelism. A window is released when any of the
following conditions is met:

\begin{itemize}
  \item \textbf{Timer expiration:} A timer starts when
  the first read enters an empty window. Subsequent
  arrivals do not reset the timer.

  \item \textbf{Count threshold:} The number of accumulated
  Read Commands reaches a configurable threshold $N$.
  Setting $N=0$ disables this trigger.

  \item \textbf{Batch Read Hint:} Acceptance of a valid
  Read Command carrying an explicit Batch Read Hint
  immediately closes the current window, including
  that command.
\end{itemize}

The interpretation of Batch boundaries follows the
project's resolution of an ambiguity in Section 5.3.2.2
of the OCP specification. Specifically, the simulator
interprets the sequence \texttt{RRBBR} as three windows:
\texttt{[RRB]}, \texttt{[B]}, and \texttt{[R]}. Thus, an
accepted Batch-hinted read closes the current window even
if it is resolved through Pending DLU forwarding or returns
a missing-fragment error without becoming a NAND
candidate. A command that is not accepted does not
affect the window state.

Window release grants scheduling eligibility rather
than imposing a barrier requiring all members to execute
or complete together. Each released read is considered
independently and remains subject to Bank availability,
Cache Buffer capacity, and ordering constraints. Reads targeting
independent Banks can therefore proceed in parallel,
while same-Bank Sense operations preserve their required
order. Conversely, a read whose page becomes a Cache hit satisfying validity and ordering checks can leave the window before its release,
without waiting for the remaining members.

\subsection{Scheduling Policies and Response Arbitration}
\label{sec:scheduling}

\begin{table}[t]
    \centering
    \caption{Scheduling policy interfaces and implemented choices.}
    \label{tab:policy-space}
    \small
    \setlength{\tabcolsep}{4pt}
    \begin{tabular}{p{0.20\columnwidth} p{0.36\columnwidth} p{0.34\columnwidth}}
        \hline
        \textbf{Policy} & \textbf{Decision} & \textbf{Implemented choices} \\
        \hline
        P1 Host Sending
        & Host command selection
        & FCFS, fill-open \\
        \hline
        P2 Window Release
        & Read-window release conditions
        & Independent triggers: timer, count threshold, Batch Read Hint \\
        \hline
        P3 NAND Scheduling
        & NAND operation selection
        & FCFS, read-priority \\
        \hline
        P4 Response Arbitration
        & Return-result selection
        & FIFO \\
        \hline
    \end{tabular}
\end{table}

HBF-EventLab separates four policy interfaces (Table~\ref{tab:policy-space}). An \emph{eligible} candidate satisfies the applicable dependency, ordering, and resource constraints. Policies rank candidates or release read windows without changing these constraints.

\begin{figure}[t]
  \centering
  \includegraphics[width=0.7\columnwidth]
    {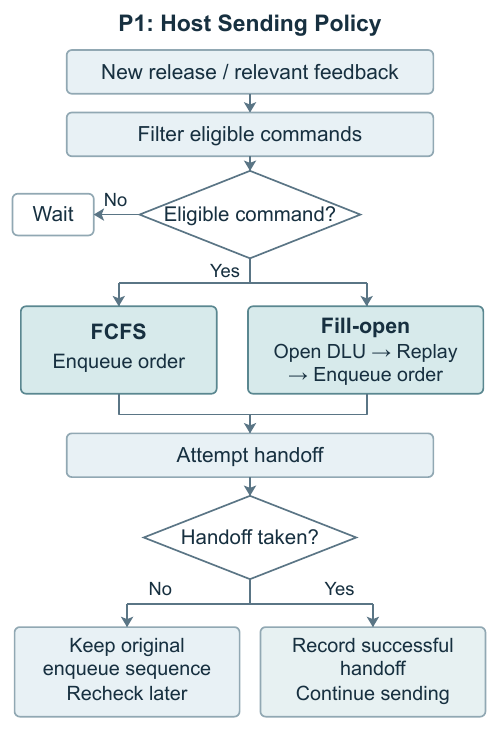}
  \caption{P1 Host Sending Policy. FCFS and fill-open select
  from the same eligible-command set and share a handoff path.
  Fill-open prioritizes commands for open DLUs, replay commands,
  and then enqueue order.}
  \Description{New workload releases or relevant feedback trigger
  the Host to filter eligible commands. If no command is eligible,
  the Host waits. Otherwise, FCFS selects by enqueue order, while
  fill-open prioritizes commands for open DLUs, followed by replay
  commands and then enqueue order. The selected command is
  handed off. If the handoff is not taken, its original enqueue
  sequence is retained for later reconsideration. If the handoff
  succeeds, the attempt is recorded and the Host continues
  sending commands.}
  \label{fig:design-host-policy}
\end{figure}

{\bf Host Sending Policy (P1).}
P1 (Figure~\ref{fig:design-host-policy}) selects the next eligible command from each channel's send queue when input or device feedback enables progress. Eligibility includes release time, dependencies, write order, recovery state, and Host-open DLU capacity; blocked candidates may be bypassed.
The current implementation provides two selection policies:

\begin{itemize}
    \item \textbf{FCFS} selects the eligible candidate with the earliest enqueue sequence, using a stable identifier to break ties.
    \item \textbf{Fill-open} prioritizes commands that supplement an already open DLU, followed by replay commands, and then enqueue sequence and stable identity.
\end{itemize}

Fill-open uses only Host-visible handoff and response records.
A rejected handoff retains its enqueue sequence; an error retry receives a new sequence. P1 can make Program or Auto Erase eligible earlier by changing fragment arrival order, but does not directly select or preempt NAND operations.

\begin{figure}[t]
  \centering
  \includegraphics[width=0.7\columnwidth]
    {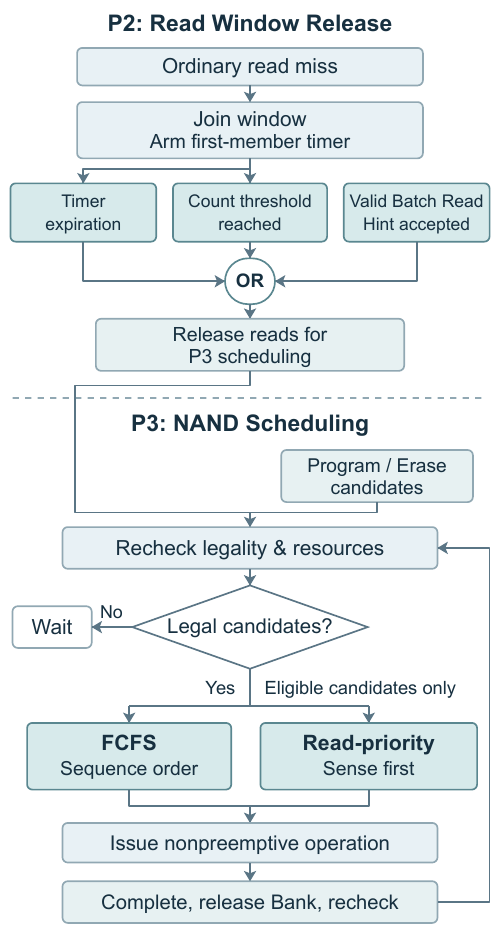}
  \caption{P2 read-window release and P3 NAND scheduling.
  Timer expiration, count-threshold attainment, and acceptance
  of a valid Batch Read Hint independently release reads for P3.
  Program and Erase candidates enter P3 independently of P2.}
  \Description{An ordinary read miss joins the Read Accumulation
  Window and arms a timer when it is the first member. The window
  is released when any one of three independent triggers occurs:
  timer expiration, reaching the configured count threshold,
  or acceptance of a valid Batch Read Hint. Released reads become
  eligible for P3 scheduling. Program and Erase operations enter
  the P3 candidate set independently of P2 release. P3 rechecks
  operation legality and resource availability, waits if no legal
  candidates exist, and selects among eligible candidates using
  FCFS or read-priority. The selected operation executes
  nonpreemptively. Upon completion, the Bank is released and
  the scheduler rechecks the candidate set.}
  \label{fig:design-read-policies}
\end{figure}

{\bf Read Window Release Policy (P2).}
P2 exposes the timer, count-threshold, and Batch-triggered release rules defined in Section~\ref{subsec:design-read}; released misses become candidates for P3 (Figure~\ref{fig:design-read-policies}).

{\bf NAND Scheduling Policy (P3).}
P3 selects among eligible Sense, Program, and Erase operations. Sense candidates require window release, an available Bank, and a Cache slot; Program and Erase candidates satisfy their respective write dependencies independently of P2.
The current implementation provides two NAND scheduling
policies:

\begin{itemize}
    \item \textbf{FCFS} selects the eligible operation with the earliest candidate sequence, using a stable identifier to break ties.
    \item \textbf{Read-priority} ranks eligible read-sense operations ahead of program and erase operations, using the same tie-breaking rules within each priority class.
\end{itemize}

Both policies are non-preemptive and preserve same-Bank sense order. Read-priority may starve writes under sustained read traffic.

{\bf UCIe Response Arbitration (P4).}
P4 selects responses staged in bounded return storage after delivery-order and capacity checks. Responses include read data, write completion, and errors.

FIFO arbitration selects the response with the earliest command-acceptance sequence among those currently eligible for transfer, breaking ties by stable identity. An unfinished earlier command does not by itself block other eligible responses. The selected response undergoes return-link
serialization and propagation before delivery to the host.
Once a read result transfers ownership to the bounded return
storage, its protected Cache Buffer reference can be released.
The return-buffer entry and corresponding command credit
remain occupied until the response is delivered.

P4 is distinct from NAND scheduling: P3 determines which
media operation occupies a Bank, whereas P4 determines
which completed result occupies the return path. Consequently,
when return storage is full, results retain their protected Cache slots, and return-link congestion can restrict subsequent read-miss issue. We keep P4 fixed to FIFO in the reported policy
experiments, isolating the effects of P1--P3 without introducing
additional response-arbitration variations.

\subsection{Event Execution and Observability}
\label{sec:event-execution}

{\bf Shared Event Queue and Causal Ordering.}
HBF-EventLab uses a shared discrete-event queue to coordinate the host, transport links, and device. Simulation time advances in discrete Ticks, with one Tick corresponding to 1~ps in the current configuration. Rather than advancing through fixed time steps, the simulator jumps directly to the next scheduled event. Events are ordered by Tick, causal round, phase, and event ID. If an event schedules another event at the same Tick with an equal or earlier phase, the new event is placed in the next causal round, preserving the causal order of events occurring at the same simulated time.

\begin{figure*}[t]
  \centering
 \includegraphics[width=0.7\textwidth]{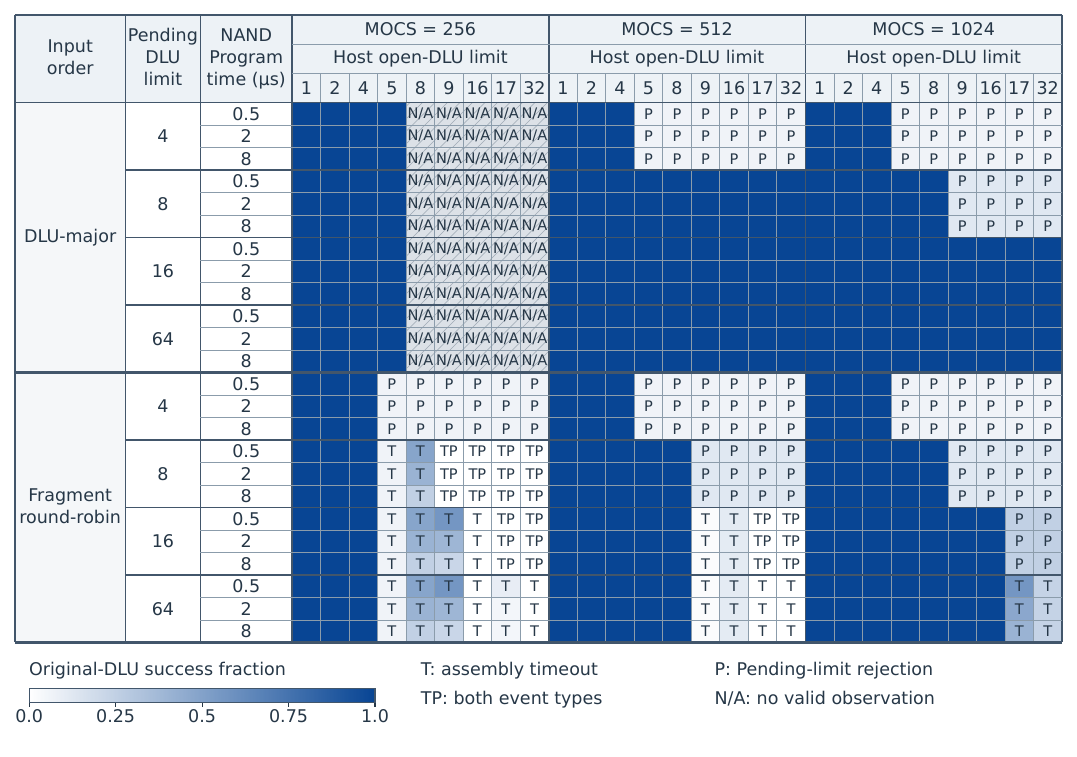}
  \caption{Write completion across 648 configurations with FCFS host scheduling. Cell color indicates the fraction of the 64 original DLUs whose fragments all complete successfully. Hatched N/A cells denote runs terminated after exceeding the 120-s wall-clock execution limit.}
    \label{fig:capacity-completion}
\end{figure*}

Assembly deadlines receive additional boundary handling. All ordinary events at a Tick are processed before the deadline batch for that Tick is frozen and evaluated. Consequently, if the final missing fragment arrives exactly at the deadline Tick, the DLU can become Ready before the timeout is processed.

{\bf Resource Completion and Event-Driven Progress.}
Completion events update component state and release resources according to their lifetimes. Targeted notifications wake affected waiters for eligibility rechecks, avoiding periodic polling.

{\bf Run Termination and Logical Outcomes.}
Run termination distinguishes normal drain, stalled unfinished work, and execution limits; unsupported inputs and internal exceptions are reported separately. Logical success or failure is recorded independently, so a drained run may still contain failed requests.

{\bf Transition Tracing and Statistics.}
Components record timestamps at state transitions, including release, handoff, acceptance, DLU readiness, window release, NAND issue/completion, and response delivery. Separate identities for logical requests, command attempts, and assembly generations preserve lifecycle distinctions across retries. Observers collect these records without modifying execution; offline analysis derives throughput and latency distributions.

\section{Backend Validation and Host--Device Coordination Analysis}
\label{sec:evaluation-backend}





We evaluate the functional correctness of the HBF-EventLab backend through controlled experiments on Ubuntu using synthetic workloads. These experiments examine how capacity and concurrency limits, address interleaving granularity, fragment release timing, and scheduling policies affect request latency and completion. The simulator models a subset of OCP HBF v0.7.0 using configurable timing parameters that have not been calibrated against hardware measurements. Experiment-specific configurations and workloads are detailed in the corresponding subsections. 
All reported timing metrics are derived from simulated timestamps, with HBF initialization excluded.

\subsection{Write Completion under Capacity and Concurrency Limits}
\label{sec:eval-write-completion}

We examine how capacity and concurrency limits, fragment input order, and NAND program time affect write completion.
Each run attempts to write 64 Device Logical Units (DLUs), each consisting of 64 fragments of 64~B. Each DLU therefore contains 4~KiB, which match the NAND page size specified in OCP HBF v0.7.0~\cite{ocp2026hbf}.
The 64 DLUs target page~0 of 64 distinct blocks within a single channel and Bank. Under the OCP specification, writing page~0 triggers an erase of the entire block~\cite{ocp2026hbf}; we set the erase time to 1~$\mu$s per block.

All fragments become available to the host at simulation time zero. We keep the host scheduling policy fixed to FCFS and compare two input orders.
\emph{DLU-major} places all fragments of each DLU consecutively. \emph{Fragment round-robin} lists the first fragment of every DLU, then the second fragment of every DLU, and so on.

The experiment varies three limits on capacity and concurrency. 
The selected MOCS value limits the number of accepted commands whose responses have not yet been delivered to the host, with each fragment consuming one command credit. The Pending DLU limit bounds the number of DLUs tracked by the device while collecting fragments, waiting for programming, or undergoing programming. The host open-DLU limit bounds the number of DLUs for which the host has begun issuing commands but has not yet received all acknowledgments.
We vary these three limits and NAND program time across both input orders, yielding 648 configurations.
The fragment assembly timeout is set to 10~$\mu$s. Automatic retries are disabled for rejections caused by the Pending DLU limit and for assembly timeouts. To bound simulation runtime, runs exceeding 120~s of wall-clock time are terminated, and their results are recorded as N/A.

Figure~\ref{fig:capacity-completion} reports the fraction of the 64 original DLUs whose fragments all complete successfully. 
Failed, cancelled, and unsent DLUs remain in the denominator. The annotations T and P indicate assembly timeout and rejection due to the Pending DLU limit, respectively; TP indicates that both occur during the run. 
A total of 60 runs are marked N/A.

\textbf{Takeaway 1: When the input order is \emph{fragment round-robin}, the MOCS limit can be reached before any DLU is fully assembled.}
With a sufficiently large Pending DLU limit, the largest tested host open-DLU limits preserving full completion under \emph{fragment round-robin} input are 4, 8, and 16 for MOCS values of 256, 512, and 1024, respectively. These observed boundaries equal the MOCS value divided by 64, the number of fragments per DLU. With larger host open-DLU limits, accepted commands can be distributed across multiple DLUs that have not yet received all their fragments. For example, with MOCS~256, a host open-DLU limit of five, and a Pending DLU limit of 64, the five DLUs contain 52, 51, 51, 51, and 51 accepted fragments when the MOCS limit is first reached. None has all 64 fragments required to begin programming, and no additional fragment can be accepted under the MOCS limit. An incompletely assembled DLU eventually times out even though the Pending DLU limit has not been reached. The corresponding commands cease to count against MOCS as their error responses reach the host.


\textbf{Takeaway 2: The input order determines how the three limits interact.}
When the input order is \emph{DLU-major}, the fragments of each DLU are grouped together, allowing assembly to finish before the device accepts fragments from many other DLUs. For example, with MOCS~512 and a Pending DLU limit of eight, all DLUs complete successfully under \emph{DLU-major} input at every tested host open-DLU limit and NAND program time. Under \emph{fragment round-robin} input, however, rejection due to the Pending DLU limit occurs when the host open-DLU limit exceeds eight. Increasing MOCS does not necessarily improve completion. With \emph{DLU-major} input, a Pending DLU limit of eight, and a host open-DLU limit of nine, increasing MOCS from 512 to 1024 introduces rejection due to the Pending DLU limit and reduces the percentage of successfully completed DLUs. A larger MOCS value permits the device to accept fragments from additional DLUs while the Pending DLU limit remains unchanged. Thus, restricting command acceptance through MOCS can help keep Pending DLU occupancy within its limit;
increasing MOCS can expose that limit.


\textbf{Takeaway 3: Longer NAND program time prolongs resource occupancy but does not remove assembly constraints.}
Across configurations with valid results, increasing NAND program time from 0.5 to 8~$\mu$s leaves the set of configurations achieving full completion unchanged, although completion percentages vary in some configurations that experience failures. Longer programming prolongs both Pending DLU occupancy and the time commands remain counted against MOCS, affecting subsequent contention. However, a fully assembled DLU does not incur an assembly timeout while waiting for or undergoing programming. These results show that write completion depends on the joint effects of the host open-DLU limit, input order, MOCS, and the Pending DLU limit, rather than on any single limit alone.


\subsection{Interleaving Granularity and Fragment Release Timing}
\label{sec:eval-interleave-arrival}


We examine how address interleaving granularity and fragment release timing affect write completion using four channels, each with one die and one Bank. The 16~KiB and 128~KiB workloads write one DLU and one complete eight-page block per channel, respectively; the latter exposes dependencies between successive page writes.

Each 64~B input fragment $i$ has Global address $64i$ and becomes available to the host at simulation time $i\Delta$, where $\Delta$ is the fragment release interval. We vary interleaving granularity over 64, 256, 1024, and 4096~B and $\Delta$ over 0, 1, 4, 16, 32, 50, 128, and 160~ns, yielding 64 configurations across both workloads.

\begin{figure}[t]
  \centering
  \includegraphics[width=\linewidth]{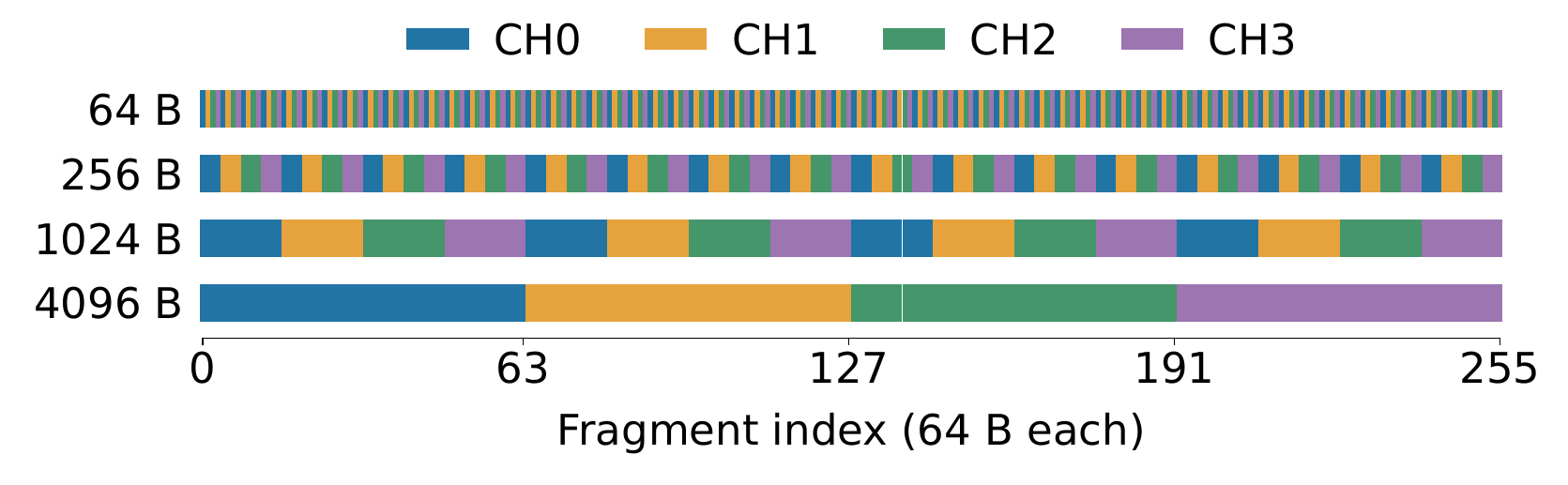}
  \caption{Planned channel mapping of the first 256 consecutive 64~B input fragments under different interleaving granularities. Rows indicate the interleaving granularity, colors identify target channels, and the horizontal axis shows the fragment index $i$, with scheduled host release time $i\Delta$.}
    \label{fig:interleave-mapping}
\end{figure}

For a fixed workload size and $\Delta$, only the mapping from Global addresses to channels and local addresses changes; the input trace and each channel's planned data volume and local address coverage remain unchanged. Figure~\ref{fig:interleave-mapping} illustrates this mapping for the first 256 fragments, with each channel's 64 fragments forming one local DLU. For this aligned sequential input, let $C=4$ and $k=I/(64\,\mathrm{B})$, where $I$ is the interleaving granularity. The scheduled release span between the first and last fragments of a DLU is
\begin{equation}
    S_{\mathrm{release}}
    = \left[64C-(C-1)k-1\right]\Delta.
\end{equation}
For the first group of DLUs, the first fragment assigned to channel $c\in\{0,1,2,3\}$ is released at $ck\Delta$. Thus, for $\Delta>0$, larger interleaving granularities shorten each DLU's release span while increasing the start-time offsets between channels.

These host release times differ from device-side assembly timestamps. With $\Delta=0$, all fragments are available at time zero, but delivery still incurs host scheduling, link transmission, and receive-side waiting. The assembly timer starts when the device creates a Pending DLU entry and applies until all 64 fragments have arrived. Erase can overlap fragment collection, while programming requires complete data, a completed erase, and satisfaction of page-order requirements. A fully assembled DLU does not incur an assembly timeout while waiting for or undergoing programming.

We use FCFS host scheduling, per-channel MOCS and Pending DLU limits of 1024 and 16, and a host open-DLU limit of 16. The assembly timeout, NAND erase time, and NAND program time are 10, 1, and 0.5~$\mu$s, respectively. Retries, replay, and fault injection are disabled. Across all configurations, per-channel peaks do not exceed 512 commands counted against MOCS and eight Pending DLUs. Both remain below their respective limits, and no rejection due to the Pending DLU limit occurs.

\begin{figure}[t]
  \centering
  \includegraphics[width=\linewidth]{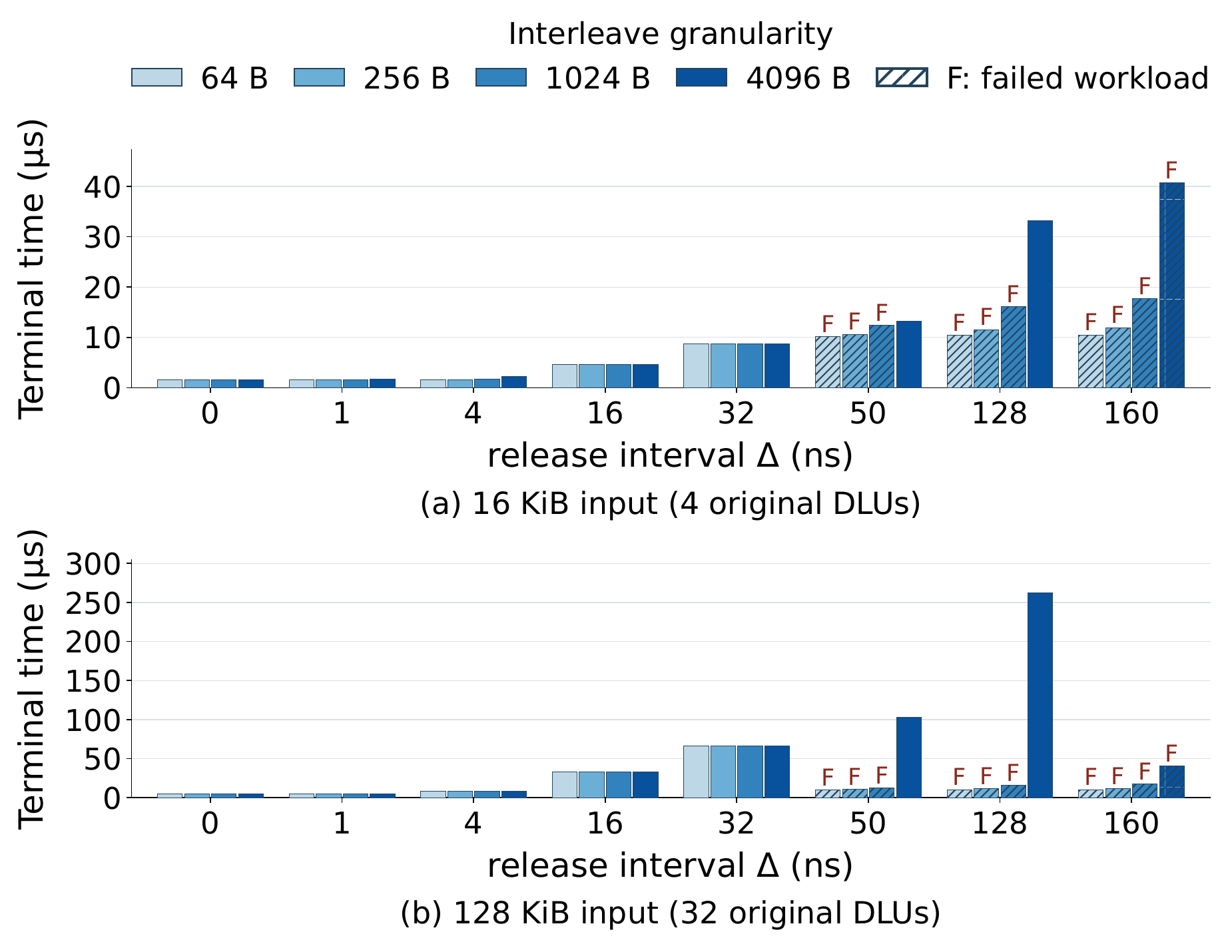}
  \caption{Workload terminal time versus fragment release interval $\Delta$ under different interleaving granularities. Terminal time is measured from the first scheduled fragment release until every original logical fragment has either completed successfully, failed, or been cancelled. The two panels use different vertical scales. With retries disabled, assembly timeouts cause the host to cancel subsequent work for the affected blocks.}
    \label{fig:interleave-terminal-time}
\end{figure}

\textbf{Takeaway 4: Larger interleaving granularities tolerate longer fragment release intervals.}
Both workload sizes exhibit the same success and failure pattern, as shown in Figure~\ref{fig:interleave-terminal-time}. Larger interleaving granularities shorten each DLU's fragment release span, allowing successful assembly at longer fragment release intervals.



\textbf{Takeaway 5: Shorter fragment release spans do not necessarily reduce workload completion time.}
Larger interleaving granularities can delay erase initiation on later channels, increasing overall completion time. When the final fragment release and subsequent programming and response delivery determine completion, differences in individual DLU release spans no longer affect completion time.

\subsection{Host Send Policies}
\label{sec:eval-host-send}



We compare the host send policies \emph{FCFS} and \emph{fill-open} to examine their effects on DLU assembly and NAND execution. Both use identical addresses, data, fragment release times, input order, and device configurations. The workload writes two 4~KiB DLUs, A and B, to page~0 of two distinct blocks in the same Bank. Each DLU consists of 64 fragments of 64~B. We use one channel, MOCS~256, a Pending DLU limit of eight, and a host open-DLU limit of two. The assembly timeout, NAND erase time, and NAND program time are 10, 1, and 0.5~$\mu$s, respectively. Retries and fault injection are disabled, and NAND scheduling remains FCFS throughout.

Both host policies use the release schedule in Table~\ref{tab:host-send-release}. At 2~$\mu$s, A has already received its first 32 fragments. B's first half and A's remaining half become available simultaneously, with B's records appearing first in input order. The host can therefore choose between completing A's assembly and starting B.

\begin{table}[t]
    \caption{Fragment release schedule used by both host send policies.}
    \label{tab:host-send-release}
    \begin{tabular}{ccc}
        \toprule
        Input records & Release time ($\mu$s) & Fragments \\
        \midrule
        1--32    & 0 & A: 0--31  \\
        33--64   & 2 & B: 0--31  \\
        65--96   & 2 & A: 32--63 \\
        97--128  & 4 & B: 32--63 \\
        \bottomrule
    \end{tabular}
\end{table}

\begin{figure}[t]
  \centering
 \includegraphics[width=\linewidth]{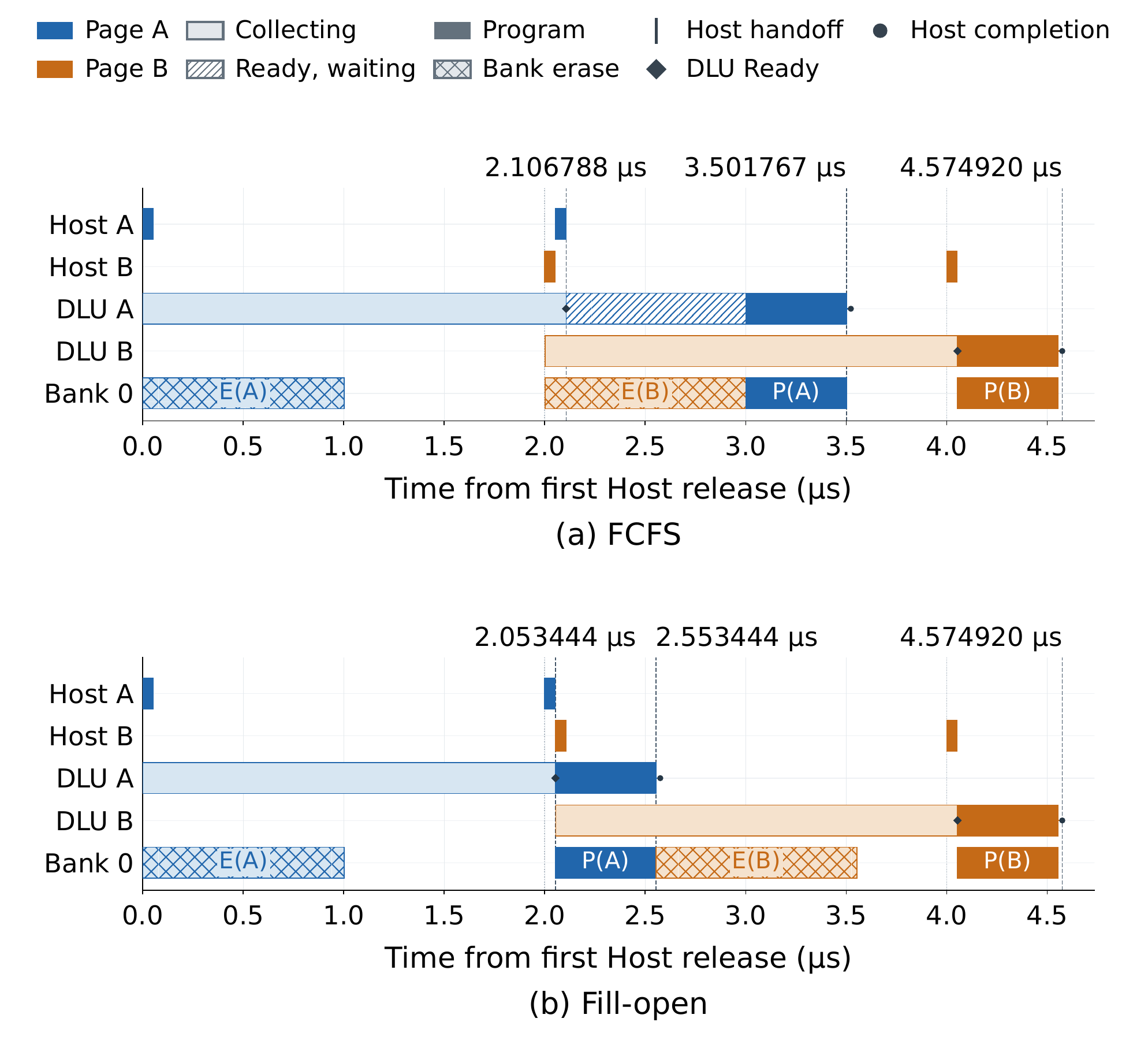}
    \caption{Host handoff, DLU assembly, and Bank execution timelines under the host send policies \emph{FCFS} and \emph{fill-open}, with NAND scheduling fixed to FCFS. Time is measured from the first host fragment release. Annotated times mark A's assembly completion, A's program completion, and workload completion. $E$ and $P$ denote block erase and page programming, respectively.}
    \label{fig:host-send-batched}
\end{figure}

\textbf{Takeaway 6: Host send policies can change NAND operation order.}
In Figure~\ref{fig:host-send-batched}, \emph{FCFS} sends B's first half before A's remaining fragments at 2~$\mu$s, so A's programming must wait for B's erase to finish. \emph{Fill-open} sends A's remaining fragments first, allowing A's programming to precede B's erase, while NAND scheduling remains FCFS. Avoiding this Bank wait advances A's final host response by approximately 0.948~$\mu$s, although its assembly completes only about 53~ns earlier.

\textbf{Takeaway 7: \emph{Fill-open} reduces mean fragment latency while workload completion time remains unchanged.}
\emph{Fill-open} reduces mean fragment latency, measured from each fragment's host release to its host completion, from 2.039 to 1.564~$\mu$s, a reduction of 23.3\%. Under both policies, B's remaining fragments are released at 4~$\mu$s, after B's erase has completed. B's remaining execution is identical, so both runs complete at approximately 4.575~$\mu$s.

\subsection{Read Dispatch and NAND Scheduling}
\label{sec:eval-read-scheduling}

\begin{figure}[h]
  \centering
 \includegraphics[width=\linewidth]{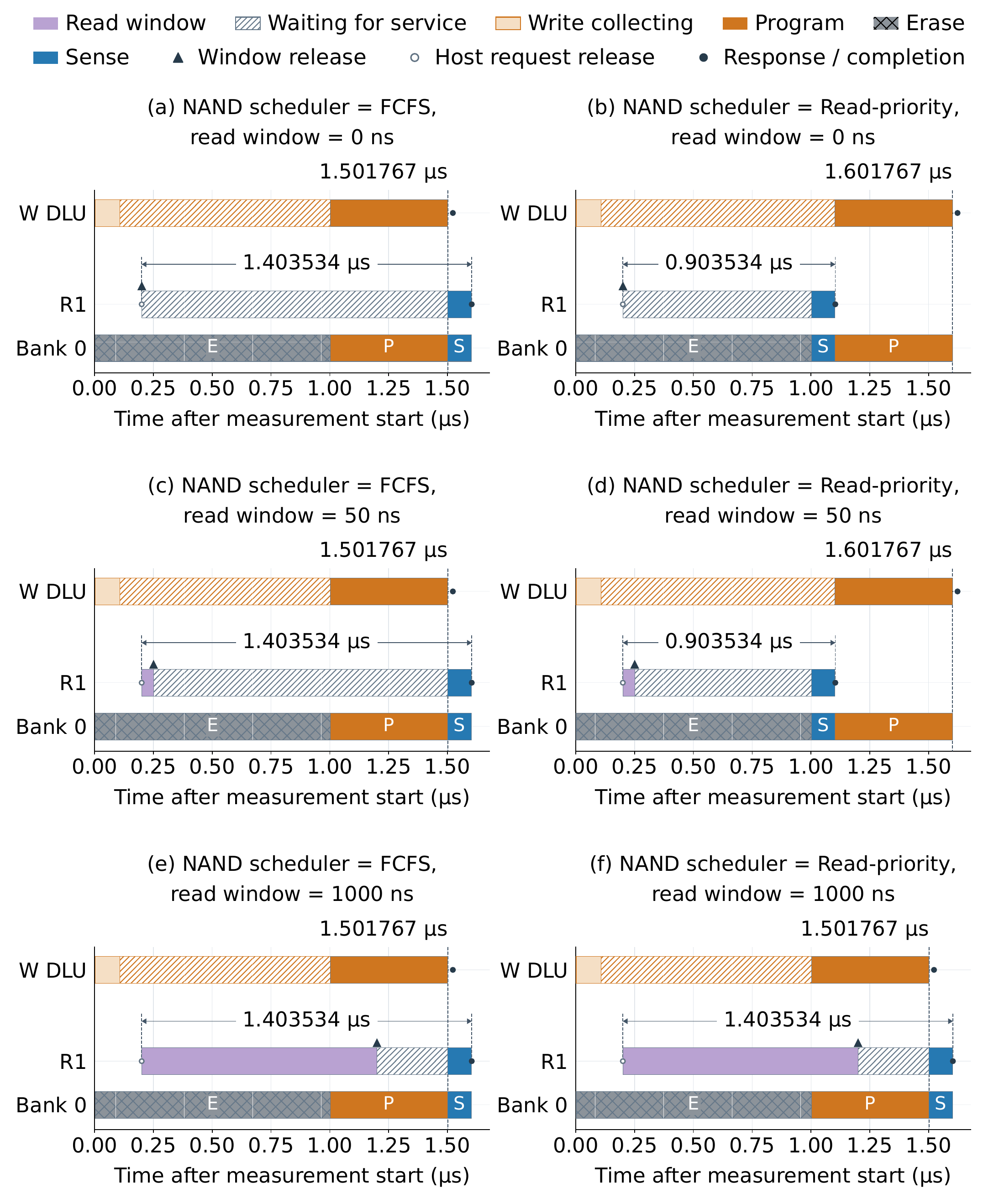}
  \caption{Read--write contention under different NAND scheduling policies and read-window waiting times. The $W$ DLU, R1, and Bank~0 rows show write progress, read progress, and Bank operations, respectively. Numbers above the $W$ DLU rows mark write program end times relative to measurement start; horizontal double-headed arrows indicate read latency from host release to host completion. $E$, $P$, and $S$ denote erase, program, and sense, respectively.}
    \label{fig:read-scheduling-timeslices}
\end{figure}

We configure the NAND scheduling policy and three read-window controls: waiting time $T_w$, read-command count threshold $N$, and an explicit Batch hint. The window timer starts when the first read enters. The window is released when the timer expires, an enabled count threshold is reached, or a valid read carrying a Batch hint is accepted.

Figure~\ref{fig:read-scheduling-timeslices} compares FCFS and read-priority NAND scheduling with $T_w=0$, 50, and 1000~ns. We set $N=0$ to disable count-triggered release and use Regular hints for all reads, so window release is controlled by the timer. The experiment uses one channel and one Bank, with the host send policy fixed to FCFS and identical inputs and resource limits across comparisons.

Four pages are initialized through the normal write path before measurement begins at absolute simulation time 10~$\mu$s, corresponding to $t=0$ in the figure. The host releases a 4~KiB write $W$ to the first page of a separate block at $t=0$, followed by a 64~B read $R$ to an initialized page at $t=0.2$~$\mu$s. The requests share the Bank but have no data dependency; $R$ requires NAND sensing rather than Pending forwarding or a cache hit. NAND erase, program, and sense times are 1, 0.5, and 0.1~$\mu$s, respectively.

Read latency is measured from $R$'s host release to its host completion. Write program end times are measured relative to measurement start and exclude subsequent response delivery.

\textbf{Takeaway 8: Short read-window waits can overlap NAND erase.}
With $T_w=0$ or 50~ns, $R$ leaves the window before erase completes, allowing its sense operation to compete with $W$'s program when the Bank becomes available. FCFS selects program first, whereas read-priority selects sense first, reducing read latency by 0.5~$\mu$s while delaying write program completion by 0.1~$\mu$s. For either NAND policy, increasing $T_w$ from 0 to 50~ns leaves the result unchanged because the additional window wait overlaps erase.


\textbf{Takeaway 9: Late window release can prevent read priority from taking effect.}
With $T_w=1000$~ns, $R$ leaves the window after $W$'s programming has started. The program operation cannot be preempted, so both policies execute erase, program, and sense in the same order and produce identical read latency.


\textbf{Takeaway 10: Count thresholds enable early release, and explicit Batch hints release the window immediately.}
In addition to timer-based release, the simulator supports a configurable read-command count threshold and explicit Batch hints, with the corresponding experiments not shown in the figure.



\section{Real-GPU Integration and Modeled Inference Timing}
\label{sec:gpu-integration}

\subsection{Integration and Timing Model}

We connect a modified llama.cpp/ggml CUDA executor~\cite{llamacpp} to HBF-EventLab to evaluate how backend access delays propagate to token completion. The real GPU performs model computation, while the backend tracks symbolic content and resource state for selected weights. TTFT and TPOT combine frozen GPU activity durations with online HBF access delays on a simulated timeline.

We first run native GPU inference and use CUPTI~\cite{cupti} to record kernel, memcpy, and memset activities within synchronized windows. Each window duration is the length of the union of its activity intervals, counting overlaps once and excluding idle gaps. We freeze the window order and durations for all compared configurations. Simulator wall-clock execution time is recorded separately.

Before computation consumes a selected weight matrix, the adapter incrementally submits reads through \texttt{TensorAccessBridge}, using the existing Host, Device, and event paths. Computation proceeds after the parent access succeeds. Parent accesses execute sequentially, while their child requests may progress concurrently. Actual weight bytes remain on the GPU; the backend models matrix-level synchronous accesses, without overlap between a selected read and its consuming computation.

Let $c_j$ be the frozen activity duration of window $j$, $\mathcal{C}_k$ the windows executed before token $k$, and $\mathcal{A}_k$ the HBF accesses completed before it. Let $w_a$ be the simulated interval from initial parent-access submission to final completion, including input backpressure, queueing, media or Cache service, and response delivery. It is measured at the parent boundary, not summed over child latencies. With initialization excluded and nonoverlapping parent-access intervals, the timelines are
\begin{align}
T_k^{\mathrm{base}} &= \sum_{j\in\mathcal{C}_k} c_j, \\
T_k^{\mathrm{int}} &= \sum_{j\in\mathcal{C}_k} c_j + \sum_{a\in\mathcal{A}_k} w_a.
\end{align}
For the eight output tokens, each timeline yields
\begin{equation}
\mathrm{TTFT}=T_1,\qquad \mathrm{TPOT}=\frac{T_8-T_1}{7}.
\end{equation}
TPOT is the mean of the seven subsequent inter-token intervals.

\subsection{Workload and Compared Configurations}

We run Qwen3-0.6B~\cite{qwen3_06b} with F16 weights on Windows x64 and an NVIDIA GeForce RTX 5070, using a fixed 44-token prompt and greedy decoding to generate eight tokens. The experiment uses one request and one sequence; the KV cache remains in native GPU F16 form.

F1 selects the FFN gate, up, and down matrices in layer 10; F2 selects layers 0, 10, 20, and 27; F3 selects all 28 layers. They register 18, 72, and 504\,MiB, respectively. Attention, embedding, and output-layer weights remain outside the HBF-selected regions. All configurations share the frozen GPU activity-time table.

The backend reference configuration, \texttt{H0-512MiB.ini}, provides 512\,MiB across one channel, one die, and two Banks, each with two 4-KiB Cache slots. It selects 256 command credits, four Pending contexts, and two Host-open DLUs. UCIe Grade~1 supplies 48\,GB/s per direction; Sense, Program, and Erase take 0.1, 0.5, and 1\,$\mu$s, respectively. Host and NAND scheduling use FCFS, response arbitration uses FIFO, and the read window has a 10-ns timer and a 32-command threshold. These are functional, uncalibrated timing settings. The repository configuration records the remaining buffer, link-overhead, and Host parameters.

\subsection{Results and Takeaways}

\textbf{Takeaway 11: HBF integration preserves numerical outputs and completes all modeled accesses.}
For F1--F3, complete logits and output tokens are bitwise identical to native GPU execution. Online checks verify selected-region coverage and logical-fragment completion, with each parent completing at the latest final completion time of its children. All runs finish without logical-access failures or command-attempt errors, and Host and Device drain normally. F3 completes 4,032\,MiB of reads through 1,032,192 Global requests and 66,060,288 fragments.

\begin{figure}[t]
  \centering
  \includegraphics[width=0.7\columnwidth]
    {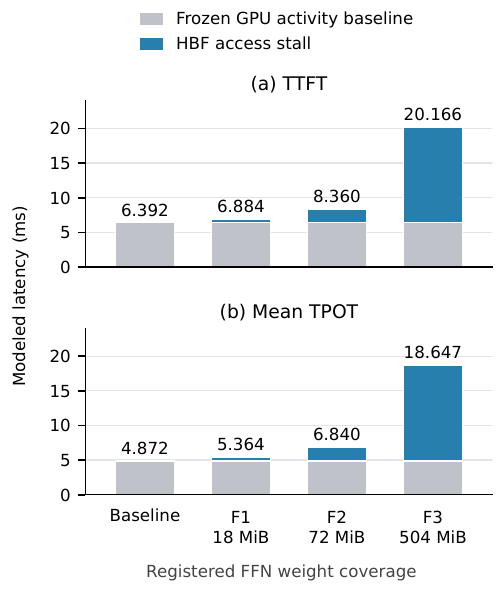}
  \caption{Modeled TTFT and mean TPOT.
  Gray segments show the frozen GPU activity baseline;
  blue segments show the HBF wait increment.
  F1--F3 cover 18, 72, and 504\,MiB of FFN weights.
  Labels give total latency in milliseconds.}
  \Description{Two vertically arranged stacked bar charts
  compare the zero-wait baseline with F1, F2, and F3.
  The upper panel shows TTFT and the lower panel shows
  mean TPOT. Gray GPU activity components remain constant
  within each panel, while blue HBF waiting components
  increase with weight coverage.}
  \label{fig:gpu-ffn-token-latency}
\end{figure}

\textbf{Takeaway 12: Broader FFN weight coverage increases modeled TTFT and TPOT.}
Figure~\ref{fig:gpu-ffn-token-latency} shows TTFT increasing from the 6.392-ms zero-wait baseline to 20.166\,ms for F3, while mean TPOT increases from 4.872 to 18.647\,ms. F1, F2, and F3 accumulate 3.936, 15.742, and 110.197\,ms of HBF waiting across eight tokens. Under this configuration and synchronous-access model, the one-, four-, and 28-layer selections produce approximately $1{:}4{:}28$ cumulative waits.

\textbf{Takeaway 13: Online HBF access delays are consistently reflected in token completion times.}
For every token in F1--F3, the integrated timestamp minus the zero-wait baseline equals the cumulative HBF waiting time, with an audited residual of 0\,ps. This verifies that access-completion feedback and token-time accounting follow the defined timing model.

\section{Discussion and Future Work}\label{sec:discussion}

{\bf Hardware calibration.}
HBF-EventLab models HBF behavior according to the OCP HBF specification, but its parameter values have not yet been calibrated against hardware measurements. Additional empirical data are needed to calibrate the model and validate its timing accuracy across different hardware configurations.

Calibration also requires identifying effects outside the current
model. HeatWatch shows that temperature, retention time, wear, and
recovery between program/erase cycles jointly affect 3D NAND
reliability~\cite{heatwatch2018}. The fixed operation durations and
symbolic contents used here support scheduling experiments; they do
not establish thermal behavior, cell error rates, or device endurance.

{\bf Integration with GPU simulators.}
The current GPU integration supports execution on a real GPU. We are extending this support to simulated GPUs in system simulators such as gem5~\cite{gem5_2011}. This ongoing work aims to enable GPU--HBF co-simulation and the study of their interactions within a unified simulation environment.

The gem5-gpu provides a precedent for coupling CPU and GPU simulation
through a shared memory-system model~\cite{gem5gpu2015}.
Physical data movement introduces further design choices: BaM
supports GPU-initiated storage accesses using a software cache and
high-throughput queues~\cite{bam2023}. Modeling such a path would
require explicit transfer, cache, and synchronization behavior beyond
the current synchronous timing adapter, which retains tensor bytes
on the GPU.

{\bf Serving workload scope.}
Application scheduling changes the requests that reach a memory
device. Orca schedules inference at iteration granularity to admit
and retire requests dynamically~\cite{orca2022}, while Sarathi-Serve
uses chunked prefills and stall-free batching to manage the
throughput--latency tradeoff~\cite{sarathiserve2024}. The current
matrix-level integration isolates how selected weight-access waits
propagate into token latency. Evaluating a serving scheduler would
add request arrivals, batching, and KV-cache traffic through the
incremental backend interface.







\section{Conclusion}
\label{sec:conclusion}

We presented HBF-EventLab, a configurable event-driven simulator for studying HBF request processing and host--device interactions. The simulator explicitly models fragmented write assembly, command admission, read dispatch, resource contention, and completion feedback. Controlled experiments show how resource budgets, address interleaving, fragment release timing, and scheduling jointly determine write completion and request latency. Integration with real GPU execution further demonstrates that selected weight-access waits can be consistently propagated into modeled TTFT and TPOT while preserving numerical outputs. HBF-EventLab provides a foundation for controlled mechanism studies, policy exploration, and future calibrated evaluation of HBF-based systems.

\bibliographystyle{ACM-Reference-Format}
\bibliography{sigconf}

\appendix









\end{document}